\documentclass[12pt]{article}
\usepackage{amsmath}
\usepackage{graphicx,psfrag,epsf}
\usepackage{enumerate}
\usepackage{natbib}
\usepackage{bbm}
\usepackage{url} 

\newcommand{\blind}{0}

\begin{document}

\def\spacingset#1{\renewcommand{\baselinestretch}%
{#1}\small\normalsize} \spacingset{1}


\if0\blind
{
  \title{\bf Non-Stationary Covariance Estimation using the Stochastic Score Approximation for Large Spatial Data}
  \author{Amanda Muyskens\hspace{.2cm}\\
    Department of Statistics, North Carolina State University\\
    and \\
   	Dr. Joseph Guinness \\
    Department of Statistics, Cornell University\\
   	and \\
    Dr. Montserrat Fuentes \\
    College of Humanities and Sciences, Virginia Commonwealth University\\
}
  \maketitle
} \fi

\if1\blind
{
  \bigskip
  \bigskip
  \bigskip
  \begin{center}
    {\LARGE\bf  Non-Stationary Covariance Estimation using the Stochastic Score Approximation for Large Spatial Data}
\end{center}
  \medskip
} \fi

\bigskip
\begin{abstract}
	We introduce computational methods that allow for effective estimation
of a flexible, parametric non-stationary spatial model when the field size is too large to compute the multivariate normal likelihood directly. In this method, the field is defined as a weighted spatially varying linear combination of a globally stationary process and locally stationary processes. Often in such a model, the difficulty in its practical use is in the definition of the boundaries for the local processes, and therefore we describe one such selection procedure that generally captures complex non-stationary relationships.
We generalize the use of stochastic approximation to the score equations for data on a partial grid in this non-stationary case and provide tools for evaluating the approximate score in $O(n\log n)$ operations and $O(n)$ storage. We perform various simulations to explore the effectiveness and speed of the proposed methods and conclude by making inference on the accumulation behavior of arsenic applied to a sand grain.
\end{abstract}

\noindent%
\vfill

\newpage
\spacingset{1.45} 
\section{Introduction}
\label{sec:intro}

 Large environmental datasets are often defined on naturally heterogeneous fields or have other inherently spatially varying conditions. Therefore, it is unreasonable to expect a response variable to be well-modeled by a stationary process over a large domain space.  However, using non-stationary models is difficult in practice due to the conceptual challenges in specifying the model and the computational challenges of fitting the model when the data is so large that memory constraints prevent formation of the covariance matrix. We propose a simple, but flexible parametric non-stationary model and corresponding computationally efficient statistical methods for estimating the model from large datasets.
 
 Our modeling approach is similar to that in \cite{fuentes2001high,fuentes2002interpolation}, but our estimation method differs and allows us to extend its practical implementation.
Fuentes models the non-stationary process $Y(s)$ at spatial location $s$ as locally-stationary processes:
 
 $$Y(s) = \sum_{i=0}^q \omega_i(s)  Z_{i}(s)$$
  where  $s$ is the spatial location and
 $Z_{i} \sim GP(\mu_i, C(\theta_i))$
 $i = 0, 1, 2....q$ and $C(\theta_i)$ is any parametric form of a covariance function. The covariance function of $Z_i$ is specified parametrically with parameters $\theta_i$, and inference on $\theta_i$ is the primary goal of estimation. The $\omega_i$ are assumed to be non-random, unknown, and positive spatially contiguous weights. Though the model specification allows the weighting functions $\omega_i$ to be quite general functions, in this paper we assume the weighting functions form a partition of the domain in order to capture non-smooth changes in the covariance. Our parameter estimation method, however, is more broadly applicable to the model and is not limited to this simplification.
 We formally define partitioning the domain $D$ by $\{ D_1,\ldots,D_q\}$ such that $D = \cup_{i=1}^{q} D_i$. Let $s \in D$ and $\omega_i(s) = 1$ if $s \in D_i$ and 0 otherwise.

 This particular formulation of the model has many application-driven advantages. First, its structure ($\omega$) has an intuitive and scientifically flexible design. For example, there may be scientifically relevant reasons to partition the field in a certain way or specific known features that are expected to influence the correlation structure of certain points. For example, in spatial data across the United States, the partition structure could be defined along state lines in order for analysis to inform policy decisions. However in the application considered in this paper, we do not have a priori knowledge of a partition for analysis. Next, the model lends itself to well-known testing and model selection procedures. Likelihood ratio tests could be performed to explore null hypotheses about the Gaussian processes in order to interpret the parameters and how they change across the field. Finally, familiar choices of covariance structures can continue to be used and interpreted as such. 
 
 In Fuentes' implementation of this model, computational necessity drives its practical application. Some adjacent points are assumed independent so that the data is divided into equally-sized blocks and analyzed individually. Similarly, \cite{risser2016nonstationary} use a partitioned model for non-stationary data, but use the covariate information to define the boundaries. In both cases, these blocks must be small enough so that the maximum likelihood parameters can be estimated directly through formation of the covariance matrix, or they must be rectangular in shape so that spectral methods such as the Whittle likelihood can be implemented. Our method of estimation relaxes these assumptions by allowing partition blocks of any size and shape as chosen by the data-driven procedure introduced in Section \ref{sec:partition}. Additionally, our model generalizes the practical application of Fuentes' model by  defining $\omega_1(s) = 1$. This defines globally stationary component that was previously computationally difficult so that all points are potentially spatially correlated \citep{fuentes2007bayesian}. 
 
 Of importance to estimating the model well is the number of processes $q$ and their accompanying structure, and we present a computationally efficient method to select this in Section \ref{sec:partition}. In \cite{risser2016nonstationary}, they define blocks only through covariate information. We instead implement a method based on likelihood ratio testing that directly uses the estimated spatial correlation of the data to cluster the locations into spatially contiguous partition blocks. To overcome previous restrictions to the model to estimate $\mu_i, \theta_i$, we propose a new estimation method involving the generalization of \cite{stein2013stochastic} stochastic score approximation to the non-stationary case. Our work describes the solving of non-linear systems with non-stationary covariances through unique application of circulant embedding, new preconditioners, and spectral density differentiation. Implementation details are provided for data in the gridded case and yield a corresponding new estimation method that is computationally $O(nlogn)$ and $O(n)$ memory.

 \cite{guinness2015likelihood} extend spectral techniques and the Whittle likelihood to the non-stationary case by using evolutionary spectral theory. They show in simulation that when a tapering method is applied, they obtain asymptotically unbiased and efficient estimates of their specified parameters. Finally, the structure selection method they present utilizes the Ising model to uncover stationary blocks of the domain. However, their estimation procedure depends on a determinant approximation that is not well understood, and since the Ising model relies on a fixed number of partition blocks, they must fully estimate parameters in several models to select a partition.   
 
Another common non-stationary model is kernel convolution \citep{Higdon1998aprocess,higdon1999nonstationary}. In this model, spatially-varying kernel functions are convolved with a stationary, often white noise, process. The parameters of the model are defined in the kernel function, and they demonstrate the model's benefits using Bayesian estimation.While this type of model is flexible, its practical implementation approximates the convolution integral with a small number of components, which leads to a low-rank covariance matrix. Our method requires no rank reduction of the covariance matrix in order to be computed quickly. 

Other classical non-stationary models have been well-studied.  Deformation models require the formation of full covariance matrices and are therefore computationally inefficient for large datasets \citep{sampson1992nonparametric}. Another attractive classical model is the moving window approach \citep{haas1995local}. However, since it involves defining the covariance with a moving window of the data, there is no guarantee that the resulting global model covariance is positive definite or can even be fully defined.

The environmental application we consider is in micro-scale soil science. When trace amounts of arsenic are dispersed into the environment, it can be harmful to life through contamination of water, plants, and soil. Although the theoretical chemical binding of pure arsenic is well understood, it is not clear how it will chemically bind in the heterogeneous conditions of the soil, where both organic matter and minerals coexist. 
By studying the micro-scale accumulation behavior of arsenic applied to a sand grain, we can characterize its spatial correlation. Studying this gives us insight to potential lurking variables that can  describe arsenic's preference to bond beyond elemental structure. However, since there are diverse elemental compositions across a 100x100 micron region, we expect the spatial correlation to vary with space. Thus, our objective is to better understand the diversity of micro-scale spatial correlation of the accumulation of arsenic on a sand grain.

In this paper, we first introduce a useful non-stationary model for large data. In Section \ref{sec:partition}, we describe a partition structure of the $\omega$ obtained from applying a new algorithm to the data, and then in Section \ref{stochasticscore} we generalize the use stochastic approximation to the score function to the non-stationary case and detail other computational tools involving the FFT for fast score computation for gridded data. Next in Section \ref{opt}, we describe an algorithm to estimate the parameters quickly. In Section \ref{sim}, we perform simulation studies to numerically validate the estimation method, and finally in Section \ref{data}, we apply our method in order to draw scientific conclusions.

\section{Partition Estimation Method}
\label{sec:partition}
\newcommand{\equivj}{\overset{\mbox{\tiny $(j)$}}{\sim}}
The first step in our estimation procedure is to estimate the partition structure of the $\omega$. These blocks define the areas of local stationarity and will be held as fixed in later parameter estimation. In some applications there may be a scientifically motivated partition of the domain. However, in many applications, the partition must be estimated from the data. Therefore, it is important to have a method for selecting possible partitions. Since enumerating all possible partitions is intractable, our strategy is to first generate an ensemble of candidate partitions and use an information criterion to choose the best partitions in the ensemble. In \cite{fuentes2001high} the partition structure is chosen via BIC from an ensemble of candidates of only equal size blocks. This structure is not likely in natural systems so we propose a method that self-selects shape and size of the partition blocks through likelihood ratio testing.

\begin{figure}
 	\begin{center}
 		\includegraphics[width=4.5in]{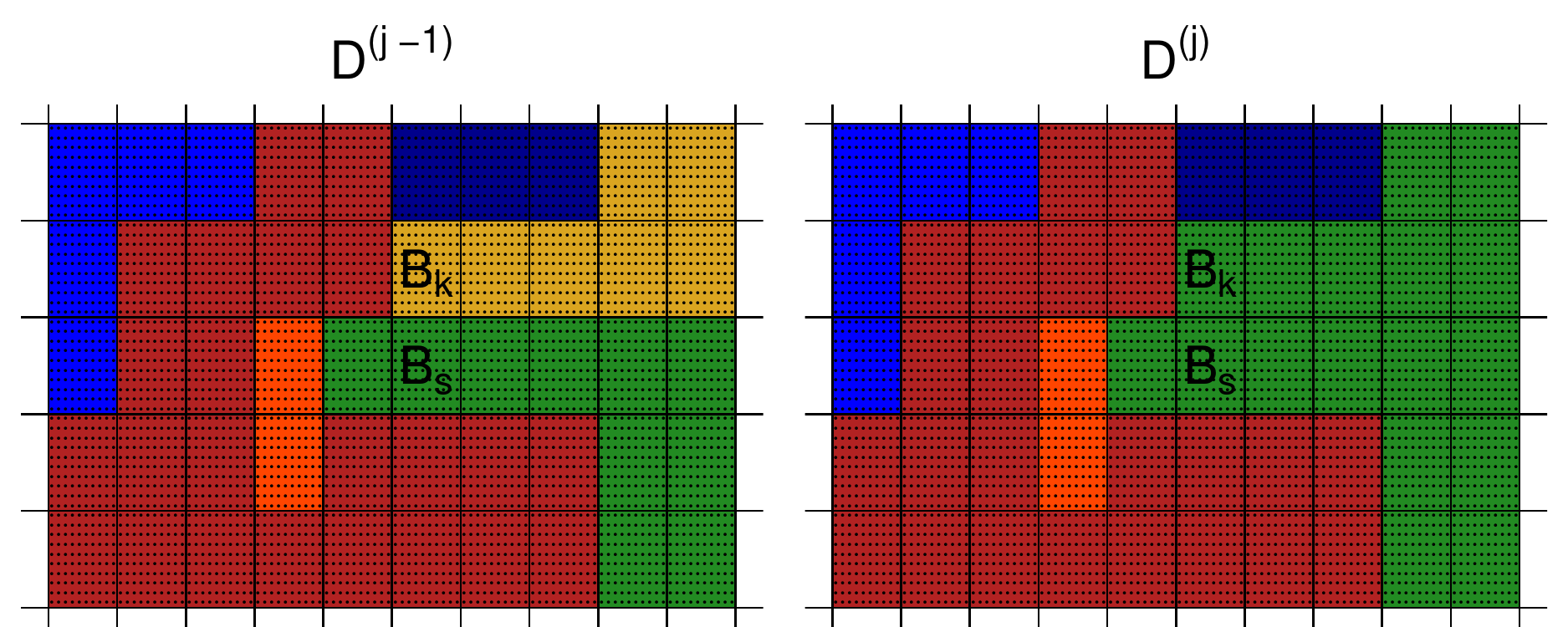}
 	\end{center}
 	\caption{Example of block joining step of $B_k$ and $B_s$ at the $j^{th}$ step of the algorithm assuming we cannot reject $H_0$ in step $j$. Color represents membership of the block to a partition segment $D_i$.}
 	\label{fig:part}
 \end{figure}

 To estimate one partition candidate, we begin by partitioning the domain into a base partition $B$.  This partition is made of $q^{(0)}$ equally-spaced blocks so that $B=\{B_1, B_2, ... B_{q^{(0)}}\}$. $B$ is chosen by the modeler to be as small as possible to estimate spatial covariance parameters, and we chose square blocks of $10 \times 10$ pixels as seen in Figure \ref{fig:part}, though other choices are possible. Our method for generating a partition $D$ is iterative and is initialized with $D^{(0)} = B$ so that $D^{(0)}_i = B_i$. At step $j+1$ of the partition selection algorithm, the partition $D^{(j+1)}$ is formed by possibly joining two neighboring blocks of the partition $D^{(j)}$, according to a likelihood ratio test to be described below. This means that any block $D_{i}^{(j)}$ consists of a subset of $\{ B_1,\ldots,B_{q^{(0)}} \}$. Partitions define an equivalence relation, in the sense that $i \equivj k$ if $B_i$ and $B_k$ are both in the same block of $D^{(j)}$. Let $\theta_i$ be a length $p$ vector of covariance parameters describing block $B_i$ and when $i \equivj k$, $\theta_i = \theta_k$.

Define the set $N_B$ of all pairs of neighbor blocks $N_B=\{ \{B_s, B_k \} | B_s$ neighbors $B_k\}$ with $n_B$ elements. Then at the $(j+1)$th step of the algorithm, we randomly sample one pair $\{B_s, B_k\} $ from $N_B$ test the hypothesis:
$$H_0: \theta_s = \theta_k$$
$$H_1: \theta_s \ne \theta_k$$
with a likelihood ratio test statistic $\Lambda(B_s, B_k| D^{(j)})$ defined below. Assume for illustration that $B_s \in D_s^{(j)}$ and $B_k \in D_k^{(j)}$. If we cannot reject $H_0$, we join the blocks setting $D_s^{(j+1)}=D_s^{(j)} \cup D_k^{(j)}$ and $D_k^{(j+1)}= \emptyset$. However, if we reject $H_0$, all values are unchanged in the update so that $D^{(j+1)} =D^{(j)}$.

The likelihood ratio test is based on the test statistic:

$$\Lambda(B_s, B_k | D^{(j)}) = \frac{\prod\limits_{i \equivj s, i \equivj k} L(\tilde{\theta}^{(j+1)}| B_i)}{\prod\limits_{g \equivj s} L(\hat{\theta}_{g}^{(j)}| B_g)
	\prod\limits_{h \equivj k} L(\hat{\theta}_{h}^{(j)}| B_h)}$$
where $\tilde{\theta}^{(j+1)}$ be the maximized parameter values such that $H_0$ is true at stage $j+1$ for the likelihood $\prod\limits_{g \equivj s} L(\theta_g| B_g)
\prod\limits_{h\equivj k} L(\theta_{h}| B_h)$ and $\hat{\theta}_{i}^{(j)}$ is the maximized parameter values for block $B_i$ at the stage $j$. We compare $-2 log(\Lambda) $ to the $\chi^2_p$ distribution to obtain a p-value for the test. This is compared to a small p-value cutoff that anticipates increased type I errors from multiplicity, and the appropriate action is made to obtain $D^{(j+1)}$.

We continue to sample neighbor pairs from $N_B$ without replacement exhaustively. The final state of the partition $D^{(n_B)}$ gives a candidate partition structure. The candidate depends on the significance level chosen for the likelihood ratio testing as well as the random sampling of the neighbor pairs. Thus, we suggest trying a variety of small p-value cutoffs and repeating this procedure to obtain a set of $l$ viable partitions $\{P^1, P^2, ..., P^l\}$. Let $\hat{\theta}^m$ be the vector of all maximized covariance parameters describing  partition $P^m$ and  $\hat{\theta}^m_i$ be the vector of length $p$ of maximized covariance parameters for data in block $B_i$ in partition $P^m$. Thus, each candidate partition has the block-independent likelihood
$$ L(\hat{\theta}^m | P^m) = \prod\limits_{i=1}^{q^{(0)}} L( \hat{\theta}^m_i| B_i, P^m)$$
We use this likelihood to calculate the BIC for each candidate $P^m$ and select the best partition from this ensemble. Simulations of the effectiveness of this approximate partition selection method can be seen in Section \ref{sim3}.

 \label{sec:sim}

\def\citeapos#1{\citeauthor{#1}'s (\citeyear{#1})}
\section{Efficient Computation of the Stochastic Score}
\label{stochasticscore}

Given the estimate of the structure $\omega$ from Section \ref{sec:partition}, we assume the definitions of the locally stationary processes are fixed. The next objective in our estimation procedure is to estimate the parameters describing the mean function ($\mu$) and the covariance function ($\theta$). Typical maximum likelihood estimation would involve an optimization algorithm such as Newton's method or gradient descent, where the score and likelihood would be evaluated repeatedly. However, this is not an option in the case of large data since memory constraints prevent formation of the covariance matrix as necessary for both evaluations. Therefore, in the next two sections we detail approximate methods where estimates of $\mu$ and $\theta$ can be obtained in large gridded datasets in a computationally efficient manner. In this section, we detail computational tools to compute a stochastic approximation to the score that will be implemented in an estimation algorithm described in Section \ref{opt}.

We extend the \cite{stein2013stochastic} stochastic score approximation to the non-stationary case. Let $\theta = (\theta_0,\ldots,\theta_{q})$ be a vector of the covariance parameters for partition $D$ and $K(\theta)$ be the covariance of the multivariate normal likelihood with $Y_0$ as the data vector with mean trend removed. We consider a mean linear in the covariates so $Y_0(s) = Y(s) - X(s)\hat{\beta}_{GLS}$. where $\hat{\beta}_{GLS} = (X^T K^{-1}(\theta)X)^{-1}X^T K^{-1}(\theta)Y$. Thus, by using the \cite{hutchinson1990stochastic} trace approximation, \cite{stein2013stochastic}  approximate the score of the multivariate normal log likelihood by:

$$S(\theta_i|Y_0) \approx \widetilde{S}(\theta_i|Y_0)= \frac{1}{2} Y_0^{T}K^{-1}(\theta)K_i(\theta) K^{-1}(\theta)Y_0 - \frac{1}{2N} \sum_{j=1}^{N} U^{T}_j K^{-1}(\theta) K_i(\theta) U_j$$
where $K_i(\theta) = \frac{\partial}{\partial \theta_{i}}K(\theta)$ and element $k$ of the $n \times 1$ $U_j$ vector is independently sampled so that $U_{jk} \sim Bernoulli(p=\frac{1}{2}; 1, -1)$, $j=1,2,3....N$. These assumptions imply that the expected value of the score is zero, making $\widetilde{S}(\theta_i|Y_0)$ a set of unbiased estimation equations. Stein et.\ al (2013) describe a dependent sampling scheme of the entries of orthogonal $U_j$ vectors. Using orthogonal vectors ensures that the trace approximation converges to the true trace as $N$ approaches $n$, but we have had good success with small $N$, in which case using orthogonal vectors provides negligible improvement. Demonstration of the comparison of the dependent sampling scheme can be seen in the Appendix A.

\cite{stein2013stochastic} considered a stationary model with data on a regular grid. This allows for the use of circulant embedding techniques to accelerate the linear solves in preconditioned conjugate gradient. In the rest of this section, we develop extensions of these methods to the non-stationary case where data is on a grid. This involves describing how to leverage circulant embedding in non-stationary models, and the development and testing of new preconditioners for non-stationary covariance matrices. Following the recommendation of \cite{guinness2017circulant}, we model the stationary covariances in the spectral domain using the quasi matern spectral density for each partitioned Gaussian Proccess $Z_i$ $i=0,1,2...,q$. With scale, range, smoothness, and nugget parameters given as respectively ($\sigma^{2}_i$, $\alpha_i$, $\nu_i$, $\tau_i$), the stationary form is given below for 2-dimensional fields:
$$f_i(\gamma)  = \sigma_i^2 [c_i (1-\tau_i) \{\alpha_i^2 + [\sin^2(\frac{\gamma_1}{2}) + \sin^2(\frac{\gamma_2}{2})]\}^{-1-\nu_i} + \tau_i],$$
where $c_i$ is the normalizing constant where and $\gamma=(\gamma_1, \gamma_2)$ are the 2-dimensional Fourier frequencies defined on the interval $[0, 2\pi ]^2$. Therefore including the overall stationary Gaussian Proccess $Z_0$, for each partition $D_i$, we induce the following stationary spectral density for each block $i$:
$$f_{0i}(\gamma) = f_0(\gamma)  + f_i(\gamma)$$

\subsection{Preconditioned Conjugate Gradient}
 \label{pcg}
 Stochastic approximation to the score is effective because it eliminates the need to form the covariance matrix, which means that instead of storing a $n \times n$ matrix, we only need to have capacity to store a $1 \times n$ vector. This is because each inversion of a matrix is adjacent to a vector $Y_0$ or $U_j$.  This means primary computational burden of computing the stochastic score itself is in the linear solves of the form $K(\theta)x = y$, where $y=Y_0$ or $U_j$. Since the covariance matrix is symmetric and positive definite, we can apply the preconditioned conjugate gradient algorithm to approximate $x$ within a threshold of accuracy, typically taken to be a vector norm of $10^{-6}$ or smaller. Preconditioned conjugate gradient is an iterative solving method computationally dominated by matrix-vector multiplication where instead of solving the linear system $K(\theta)x=y$, we solve the equivalent equations $P(\theta)K(\theta)x=P(\theta)y$ where $P(\theta)$ is an $n \times n$ preconditioning matrix. In order for $P(\theta)$ to be an effective preconditioner where it speeds up computation time, the condition number of $P(\theta)K(\theta)$ should be smaller than the condition number of $K(\theta)$, and the forward multiplication $P(\theta)x$ should be fast, ideally less than $O(n^2)$. 
 
 \cite{anitescu2012matrix} fit stationary models and explore preconditioners based on the inverse of the spectral density. We extend their work to our non-stationary covariance case by considering 4 possible preconditioners with different spectral densities $g(\gamma, s)$ at spatial location $s$. The forward multiplication of the preconditioners with a vector $x$ is of the form
 $$ \sum_{s=1}^{n}[\sum_{\gamma =1}^{n}g(\gamma, s') e^{-i \gamma (s-s')}] x(s)$$ 
 The resulting multiplication of each of these preconditioners is $O(n \log(n))$ due to circulant embedding using the Fast Fourier Transform (FFT). The forms of spectral densities we consider are
 \begin{align*}
  & g_1(\gamma, s) = \frac{1}{f_0(\gamma)}\\
  & g_2(\gamma, s) = \sum_{i=1}^q  I\{s \in D_i\} \frac{1}{f_i(\gamma)}\\
  & g_3(\gamma, s) = \frac{1}{f_0(\gamma)} + \sum_{i=1}^q  I\{s \in D_i\} \frac{1}{f_i(\gamma)}\\
  & g_4(\gamma, s) = \frac{1}{\sum_{i=1}^q f_{0i}(\gamma)}
   \end{align*}
 \cite{guinness2013transformation} explore similar forms of preconditioners under a different non-stationary model in the time series case.

\begin{figure}
	\begin{center}
		\includegraphics[width=6.5in]{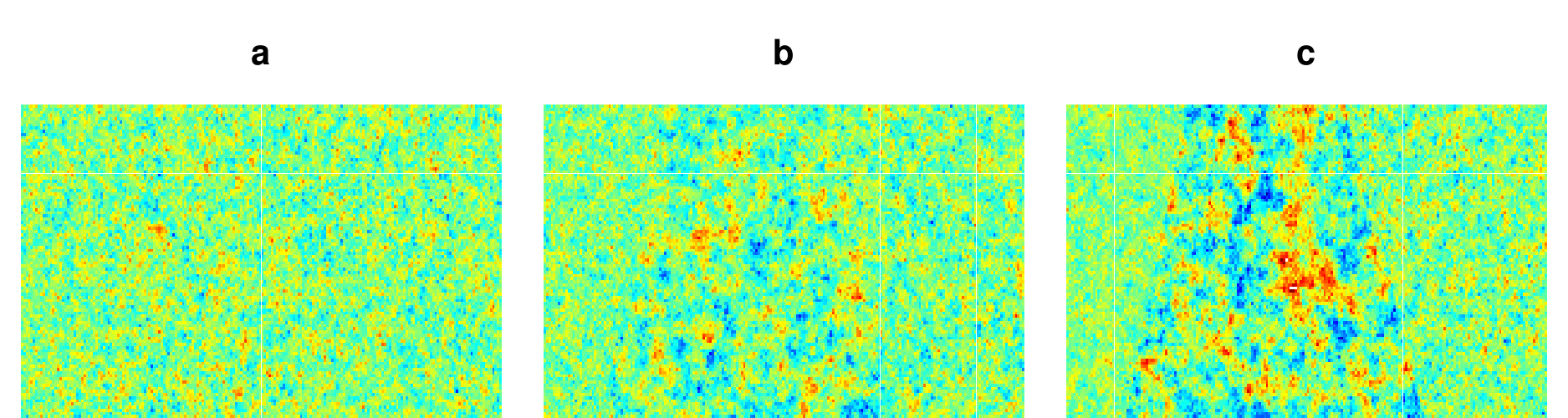}
	\end{center}
	\caption{Sample draws from 3 non-stationary models with increasingly different parameter settings. We use these parameter settings for simulations in Section \ref{sim}.}
	\label{fig:condition}
\end{figure}

To test these preconditioners, we generate parameter values from three settings with three partition blocks with sample data draws seen in Figure \ref{fig:condition}.  An ideal preconditioner would perform well in all scenarios so that it would quickly solve the linear system throughout the algorithm smoothly no matter the starting or true maximum likelihood estimates of the parameters.  We record the time to convergence within a set tolerance using the various methods for a sample size of 20,000 data points with a $200 \times 100$ data matrix for 500 replications. We give a poor starting value for all algorithms, but expect the algorithms to perform much better in practice since we use the solution from the previous step in the estimation as the starting value.  

All preconditioners were effective at significantly reducing the convergence time and number of iterations under all parameter settings.  We choose to implement the preconditioner $g_2$ in simulation since it performs well in all cases, but other choices may be selected and will only impact speed of the algorithm.
 \label{sec:sim}

%

\begin{table}[ht]
	\centering
	\begin{tabular}{|r|cc|cc|cc|}
		\hline
		& \multicolumn{2}{c|}{a} & \multicolumn{2}{c|}{b} & \multicolumn{2}{c|}{c}  \\
		\hline
		Pre-Conditioner & Time & Iters & Time & Iters & Time & Iters  \\ 
		\hline
		Not Conditioned & 0.80 (0.001)& 59 & 1.64 (0.003) & 118 & 2.29 (0.005) & 164 \\ 
		Stat ($g_1$) & 0.26 (0.000) & 14 & 0.61 (0.001) & 34 & 0.89 (0.001) & 49 \\ 
		Block Stat ($g_2$) & 0.47 (0.003) & 18& 0.57 (0.001) & 22 & 0.72 (0.004) & 28 \\ 
		Combo ($g_3$) & 0.29 (0.001) & 10 & 0.62 (0.003) & 21 & 0.88 (0.002) & 30 \\ 
		Weighted Stat ($g_4$) & 0.39 (0.001) & 16 & 1.09 (0.003) & 46 & 1.82 (0.004) & 77 \\ 
		\hline
	\end{tabular}
	\caption{Mean total time in seconds until convergence and median iterations until convergence of conjugate gradient solve using various preconditioner matrices. Convergence is defined as the square of the $L^2$-norm of the error vector less than 1e-4. Parentheses are standard error of the estimate and the standard error median iterations is less than 0.006 in all cases.}
	\label{tab:precond}
\end{table}

The procedure of matrix-vector multiplication within the preconditioned conjugate gradient is typically $O(n^2)$, but since we assume our points to be on a grid, we speed up this process by circulant embedding. In this non-stationary case, circulant embedding is applicable because we can rewrite matrix-vector multiplication as follows

\begin{align*}
\left[ K(\theta) x \right]_\ell = \sum_{j=1}^n K_0(\theta_0) x_j +
\sum_{i=1}^q w_i(s_\ell) \sum_{j=1}^n K_i(\theta_i) w_i(s_j) x_j,
\end{align*}
which can be computed with $q+1$ FFTs since each of the individual matrices $K_i$ come from stationary processes.
The above process is $O(n \log n)$, but we further accelerate computation by approximating the circulant embedding with expansion factor $\frac{5}{4}$ \citep{guinness2017circulant}. Although smaller embedding matrices are possible for parameters that belong to the locally stationary Gaussian proccesses, we choose to embed in the size of the entire field in order to increase accuracy of the approximation without adding much to the computing time or limiting shape or size of the selected partition blocks.

\subsection{Differentiated Circulant Embedding}
\label{ce}

Also vital to computing the approximate score, forwards multiplication of $K_i(\theta) = \frac{\partial}{\partial \theta_{i}}K(\theta)$ must be fast without formation of $K_i(\theta)$. It is often that covariance functions in spatial domain are slower than the spectral density to evaluate. For example, the Mat\'ern covariance includes a Bessel function in spatial domain, but only includes ordinary operations in spectral domain. We can differentiate the spectral density in order to quickly multiply $K_i(\theta)$ using circulant embedding. We include the derivatives of the quasi-matern spectral density for a stationary covariance in Appendix B for reference. Since our model induces a covariance that is the weighted sum of stationary covariances, this principle can be applied to each of the stationary covariances.

Note that if the field is not defined on a partial grid, circulant embedding is not typically an option. However, if a finer grid can well approximate to the data locations, circulant embedding could still be potentially implemented since missingness does not deteriorate its use. Alternately, methods of \cite{chen2014fast} could be implemented in order to retain $O(n log(n))$ matrix-vector multiplication.

\section{Nonlinear Solver} 
\label{opt}
Finally in this section we describe the solver with which one can ultimately obtain parameter estimates for our model. We use the methods from the previous section to compute an approximate gradient at each step. However, this problem is non-standard optimization since we cannot evaluate the likelihood to verify our the appropriateness of our step size. Because of this and a few other computational challenges specific to this problem, we design a parameter estimation algorithm as described in this section.

In parameter estimation, if some spatial covariance parameters become too large or too small, the covariance matrix is numerically singular. Therefore, we implement a proximal gradient descent algorithm on the negative log likelihood to estimate the covariance parameters $\theta$ \citep{nesterov2005smooth}. This algorithm has the same computational complexity in each step as a regular gradient descent algorithm, but includes a projection into a restricted parameter space in each step. 

Proximal gradient is a descent algorithm on a convex problem, and therefore step size selection is often straightforward. However, since we cannot evaluate the likelihood directly, there is no direct method of step size selection. Other methods derive a fixed step size, but they require deriving boundings that have no closed solution for our likelihood. In order to select a step size in each step of estimation, we pose the problem as a non-convex, non-linear solve. Hence, we make all entries of the score close to 0 as a means of estimating the parameters.  Standard non-linear solvers such as \cite{dennis1996numerical} converge slowly or not at all in our case since the score for many of the covariance parameters is also small near the boundary or outside evaluative space. For that reason, we write our own algorithm that is designed to not only make the score of each parameter close to 0, but guaranteeing it is in the interior of the parameter space. Therefore, we choose the step size so that the sign of the score is unchanged for all parameters. Since the quasi-matern parameters are often on very different scales, we leverage thresholding in order to move only the furthest parameters towards their solution first, which greatly speeds up the maximization process. Finally, as a stopping criterion, we stop progress when the maximum score value is below a threshold or the relative movement of all parameters from a step is less than 1e-6.

In order to leverage the exact closed form solution for an overall scale parameter, we reparameterize $\sigma_i^2 = \phi_i \sigma_0^2$ for $i=1,2,...q^{(0)}$ so that $K(\theta) = \sigma_0^2 \Omega(\theta)$. The likelihood for $\sigma_0^2$ is maximized by 
$$\hat{\sigma_0^2}= \frac{1}{n}Y_0^{T} \Omega(\theta)^{-1}Y_0 $$
Similarly, maximum likelihood estimates for the linear mean parameters can be explicitly solved in each iteration. Combining these tools, we are able to obtain estimates of mean and covariance parameters as needed effectively. 

\section{Simulations}
\label{sim}
We perform two simulations to explore our two-step estimation procedure. First, we assume the partition structure is known and compare our stochastic score estimation method to its primary competitor in smaller sample size test cases. Second, we test our procedure in the realistic case for data analysis where both partition and the parameters are unknown with more realistic sample sizes.
\begin{enumerate}
	 \item We assume the correct partition structure is known and compare the computation time and accuracy of our estimation method to Vecchia's Likelihood Approximation \citep{vecchia1988estimation}. We also evaluate the impact of the choice of number of vectors $N$ in the stochastic score approximation. We assume zero mean and measure accuracy in the likelihood gain of estimation vs oracle parameters.
	 \item We show use of our method with the realistic assumptions where the partition structure is unknown and the parameters are estimated. First, we extract the selected partitions and compare their success using clustering criterion. Then, we evaluate the overall effectiveness of our approximation method. We compare our method to the previously employed equal partition division. Here we also estimate a non-constant mean function as described in Section \ref{stochasticscore}.
	 \end{enumerate}
	 Through all simulations, we generate non-stationary data under our model where the form of stationary covariances is quasi-Mat\'ern. The partition setting and a sample draw from parameter settings for this generation can be seen in Figure \ref{fig:condition}. All timings were obtained using an Intel i7- 6700HQ with 16 GB of DDR3 RAM running Windows 10 in the 64 program R version 3.4.3 with Microsoft R Open.

\subsection{Known Partition Estimation Simulation}
\label{sim2}
Our primary goal in this simulation is to evaluate the accuracy and time investment of our score approximation method. Therefore we focus only on the second stage of our analysis method and assume the partition is correctly selected in order to analyze only speed and accuracy of this stage of the estimation. We compare our score approximation method to its main computational competitor: estimation via approximate likelihood using the Vecchia approximation \citep{vecchia1988estimation}.

Joint distributions can always be written as a product of an ordered sequence of conditional distributions. Vecchia's likelihood replaces the random variables in the conditioning sets with a subsets in order to reduce computational burden. Here, we consider subsets of 30 nearest neighbors,

$$ L(\theta , Y_0(s_1), Y_0(s_2),...,Y_0(s_n)|D) \approx L(\theta, Y_0(s_1)|D) \prod_{i=2}^{n} L(\theta , Y_0(s_i) | Y_0(S_i), D)$$
where $S_i$ is the set of nearest neighbors to $s_i$ in the conditioning set \citep{vecchia1988estimation}.

 We generate multivariate normal data with zero mean under the assumption that our non-stationary model is the truth. We consider varying grid sizes of $20 \times 40$, $26 \times 52$, $30 \times 60$, $36 \times 72$, and $40 \times 80$. We compare these results under 50 sample replicates assuming the underlying partition is known. This allows us to explore the accuracy of estimates and the time required to obtain them.

In addition to varying sample size, we explore how many $U_j$ vectors ($N$) are necessary for sufficient approximation to the score function. Thus for each of the settings above, we consider the score method for $N=1$, $N=5$, and $N=20$. As the number of vectors approaches the sample size, we know that the approximation to the score becomes more accurate, but the computing time is slowed. We investigate which of these settings is sufficiently accurate in order to obtain estimates as fast as possible.

As a criterion for evaluation of the accuracy, we define likelihood gain as $$ 2 \log L(\hat{\theta}, Y_0) - 2 \log L(\theta, Y_0),$$ where $L$ is the likelihood, $\theta$ are the simulation setting parameters, and $\hat{\theta}$ are the maximum likelihood parameters using approximate methods. Note that because $\theta$ oracle parameters and are not in fact the maximum likelihood parameters, there is likelihood to be gained by even approximate estimation. This leads to potentially positive likelihood gain values, and a larger value is preferred because our goal is to approximate the maximum likelihood estimates.

\begin{table}[h]
	\begin{center}
		\begin{tabular}[width=7in]{|c|c|ccccc|}
			\hline
		& Method & n=800 & n=1352 & n=1800 & n=2592 & n=3200 \\ 
			\hline
		&	Score $N=1$ & 7.60(0.63) & 12.07(0.86)& 15.23(1.04)& 19.45(1.38) & 25.67(2.20)\\ 
a		&	Score $N=5$ & 14.57(1.10)& 24.68(1.67) & 30.54(1.58) & 39.07(2.10) & 57.28(3.80)\\
		&	Score $N=20$& 45.36(2.52) & 81.82(5.44)& 96.94(4.69)& 122.95(5.20)& 173.24(10.80)\\ 
		&	Vecchia App.& 113.35(4.23)& 211.23(5.96) & 288.44(11.39) & 451.00(16.07)& 604.18(27.00)\\ 
	
		\hline
		&	Score $N=1$ & 20.38(4.97) & 36.94(4.72) & 31.93(2.21) & 62.77(8.53) & 69.80(8.44)	\\ 
	b	&	Score $N=5$ &  33.25(8.70) & 44.22(3.90) & 90.12(18.91) & 104.25(10.34) & 129.22(18.10)  \\
		&	Score $N=20$&  74.86(4.04) & 142.50(12.79) & 236.48(26.15) & 299.07(29.87) & 414.35(64.00) \\ 
		&	Vecchia App.&  111.70(4.92) & 194.88(8.22) & 263.04(10.91) & 385.56(16.56) & 476.37(13.62) \\

		\hline

				&Score $N=1$ & 25.42(5.49) & 28.65(2.89) & 51.64(11.89) & 66.03(14.13) & 98.47(21.66) \\ 
			c	&	Score $N=5$ & 28.28(2.43) & 54.28(4.09) & 65.98(3.50) & 89.30(4.26) & 133.48(9.40) \\ 
				&	Score $N=20$ & 87.17(9.03) & 171.06(23.57) & 199.84(12.36) & 276.94(16.13) & 378.64(22.04) \\ 
				&	Vecchia App& 128.53(4.68) & 227.41(8.12) & 315.01(10.81) & 465.68(16.81) & 627.62(20.42) \\ 
		\hline
		\end{tabular}
		\caption{Mean time in seconds until convergence of estimates in various methods and sample sizes in simulation. In parentheses is the simulation standard error.} 
	\end{center}
    \label{tab:vec}
\end{table}

\begin{figure}
	\begin{center}
		\includegraphics[width=6.5in]{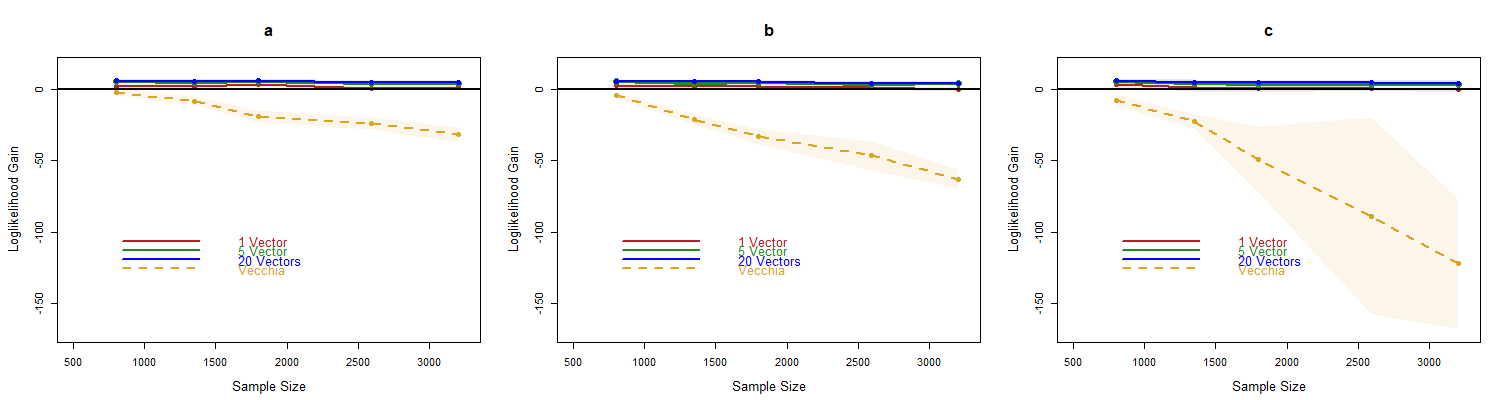}
	\end{center}
	\caption{Mean likelihood gain under non-stationary settings (a,b,c). The shaded regions represent 95\% pointwise confidence intervals for the mean value in simulation.}
	\label{fig:vecchia}
\end{figure}

As sample size increases, estimation using the Vecchia likelihood approximation is less accurate. This mean difference is greater, with a larger variance, as the parameter settings become more non-stationary. These results can be seen in Figure \ref{fig:vecchia}. Note that although the likelihood lost increases as the sample size increases, it is due to the general increase in likelihood. In fact, the likelihood lost by the method performs relatively proportional in all sample sizes. On the contrary, in all sample sizes, the stochastic score methods on average gain likelihood as compared to the oracle parameter settings. Likelihood results from $N=5$ and $N=20$ are non-statistically significantly different. In a few cases, these results are statistically more accurate than those where $N=1$. Therefore, we conclude that $N=5$ vectors is sufficient for estimation, and we include only $N=1$ and $N=5$ in the next simulation.

In terms of computation time, the stochastic score estimation method also outperforms estimation with the Vecchia likelihood in all cases for $N=1$ and $N=5$.  Even in the relatively small sample size of 3200, it takes just over 10 minutes on average for the Vecchia likelihood to converge in non-stationary setting a, but only about 25 seconds for the stochastic score approximation with 1 vector to reach a statistically better solution. This approximately 24 times speed up is significant even in small samples and is prohibitive in large samples. Since we need to perform roughly five times as many linear solves in $N=5$ as compared to $N=1$, we would expect time to scale as such. However, the computation time for the solution with $N=5$ is generally less than two times as long. This implies that each step taken with more vectors actually moves closer to the solution, and therefore less iterations are needed. Hence, we conclude that especially in larger samples, the stochastic score approximation as more accurate and faster than the Vecchia approximation at estimating the maximum likelihood parameters in our non-stationary model.

\subsection{Unknown Partition Estimation Simulation} 
\label{sim3}

Finally, we design a comprehensive simulation where we evaluate the loss in likelihood using our method in the situation where the partition structure and maximum likelihood estimates are unknown. We also allow for a non-constant, linear mean function that is simultaneously estimated with the covariance parameters. First, we fit a simple linear regression model to the data and use the residuals to estimate a partition structure for a candidate pool of 30 partition structures as previously described. As a base partition, we divide the field into small sub-blocks $B_k$ of size $10 \times 10$.  We consider the p-value cutoffs for the likelihood ratio tests as (.0005, .001, .002, .003, .004, .005, .006, .007, .008, .009). Then we select the best partition structure using BIC. We use the original simulated data to estimate the parameters using the stochastic score method with 1 and 5 vectors. For comparison, we also use the stochastic score method to estimate parameters using an equally divided partition structure with 2 blocks, as this model has previously been implemented. We explore the accuracy of the estimation over the varying sample sizes of $50 \times 100$, $70 \times 140$ and $90 \times 180$ using the parameter settings with realizations seen in \ref{fig:condition}.

This two-step estimation procedure begins with partition selection. Therefore we evaluate the success of our method through the Rand index \citep{rand1971objective}.  This criterion was conceived to evaluate the accuracy of clustering, and it is applicable here as the partition selection procedure is essentially a spatially-contiguous clustering problem.  Note however, this problem is more challenging than typical clustering because we do not assign clusters based on grouping values, but rather by the relationship among them (ie. spatial correlation). The Rand index is defined as the proportion of these points that are correctly grouped based on the true model partition $P_i$.  Mathematically, the Rand index when there are $q$ sub-blocks is 

$$Rand = \frac{1}{{n \choose 2}}\sum_{i=1}^{n} \sum_{j<i} I \{t_{ij} = \widehat{t_{ij}}\}$$

where ${n \choose 2}$ is the total number of pairs of points, $t_{ij}$ is an indicator of whether observation $i \sim j$ in the true model, and  $\widehat{t}_{ij}$ is a similar indicator for the estimated partition.

\begin{figure}
	\begin{center}
		\includegraphics[width=4in]{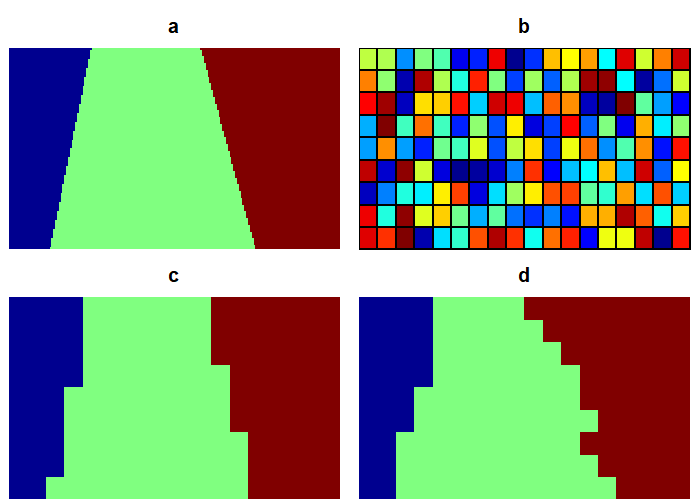}
	\end{center}
	\caption{a. True partition b. Base partition c. Best possible partition with assumed base structure (Rand=0.96) d. Minimum BIC partition from random iteration for setting 3c (Rand=0.92).}
	\label{fig:es}
\end{figure}

Although inefficient, the memory of our available hardware allows for the true evaluation of the likelihood at these sample sizes so we again use likelihood gain again as a measure of our accuracy as in the last simulation. For reference of these gain values, we also include the mean true 2 * log likelihood value using the true partition in each setting.


\begin{table}[ht]
	\centering
	\begin{tabular}{|cc|ccccc|}
		\hline
		& n & \# Blocks  & Like Gain N=1 & Like Gain N=5 & Like Gain Eq Part  & True Like \\ 
		\hline
	&	5,000 & 2.82 (0.0619) &  -66.7 (2.5) & -65.2 (2.6) & -139.1(3.0) & 3972.6 (11.8)  \\ 
	a &9,800 & 3.08 (0.0388) & -129.8(4.9) & -127.4(4.8) & -240.9(5.9) &7772.0 (17.8)\\ 
	&	16,200 & 3.08(0.0388) & -193.0(9.5) & -190.2(9.2)& -344.3 (9.0) & 12805.4 (25.8)\\ 
	\hline
	&	5,000  & 3.06 (0.0339) & -112.0(4.3) & -108.6(4.3)& -261.8(4.6) & 4860.4 (15.7)\\ 
b	&	9,800 & 3.16(0.0524)  & -188.1 (7.1)& -184.4(6.9) & -425.6(7.2) &9443.9 (22.2)\\ 
	&	16,200 & 3.26 (0.0689)  & -287.8 (12.8) & -282.9 (13.1) & -602.7 (12.5) &15554.0 (24.5)\\ 
	\hline
	&	5,000 & 3.04(0.0280)  & -152.3 (4.8) & -149.2(4.6)& -421.9 (6.6) & 5160.2 (16.1)\\ 
c	&	9,800 & 3.18 (0.0549)  & -246.7(6.6) & -243.3(6.6) & -632.2 (7.9) &10020.8 (19.1) \\ 
	&	16,200 & 3.40(0.0808)  & -364.7 (11.7) & -354.9 (10.2)& -895.6(11.9) &16574.4 (21.0) \\ 
		\hline
	\end{tabular}
\caption{Results from the unknown partition simulation that include estimated number of partition blocks, log likelihood gain as previously defined for 3 implementations of the stochastic score estimation, and finally, the true log likelihood value $2 \log L(\theta, Y_0)$, where the true simulation parameter settings $\theta$ and partition are used.}
\end{table}

\begin{figure}
	\begin{center}
		\includegraphics[width=4.5in]{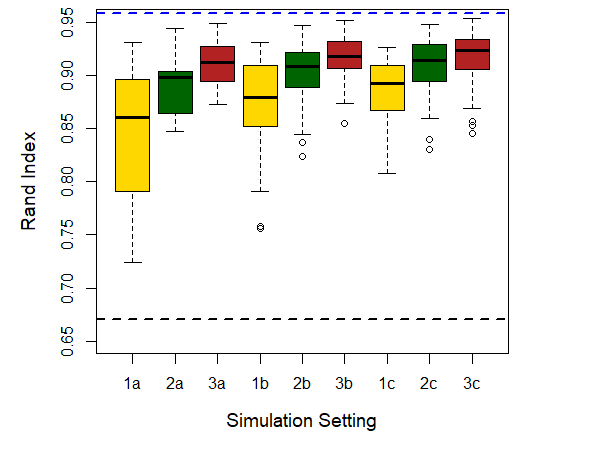}
	\end{center}
	\caption{Box plot of Rand indexes for sample sizes (1,2,3) and non-stationary settings (a,b,c). The blue line represents the maximum possible Rand index given the base partition for the largest sample size, and the black line represents the Rand index for equal block division.}
	\label{fig:randbox}
\end{figure}

The automatic partition selection methods always produce better results than the traditional approach of simply splitting the domain in half as seen in Figure \ref{fig:randbox}. So while we expect the performance of partition selection to depend on the model and p-value cutoffs chosen, we expect the methods to outperform traditional ad-hoc partitioning methods. Demonstration for the justification of selection criterion based on BIC as a proxy for the Rand index can be seen in Appendix D. Although unrestricted, the algorithm selects the correct mean number of blocks as the closest whole number in all settings. In the more non-stationary simulation settings, the partition selection method produces partitions with a higher mean Rand index with a smaller variance. This difference is settings is smaller in the larger sample sizes and therefore the method performs well in all cases with a large sample size. With the base partition we select, the closest possible partition to the truth has a maximum Rand index of 0.96. One example of a selected partition can be seen in Figure \ref{fig:es}.

\begin{figure}
	\begin{center}
		\includegraphics[width=3in]{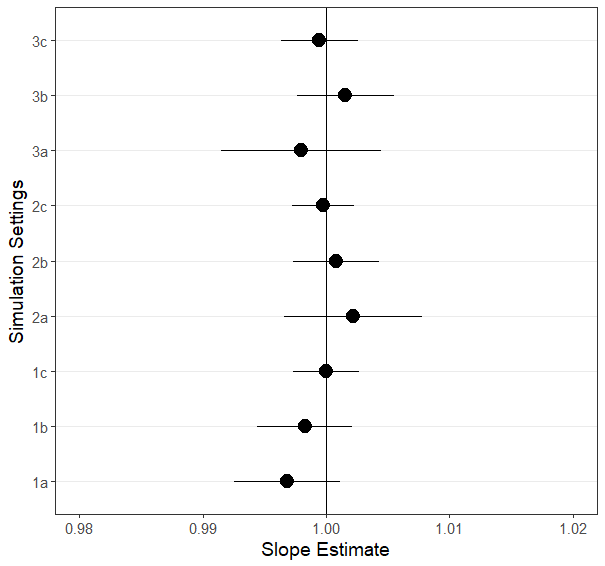}
	\end{center}
	\caption{Each plot shows simulation mean slope estimate confidence intervals for sample size (1,2,3) and non-stationary setting (a,b,c). All intervals contain the true slope parameter 1 and are therefore unbiased in simulation.}
	\label{fig:bias}
\end{figure}

Because we estimate both the partition structure and the estimates, we do lose some log likelihood as compared to the oracle parameters. However, this loss is small relative to sample size. In these larger sample size settings, there is no statistical difference in the log likelihood value lost by the stochastic score method using $N=1$ and $N=5$. However, there is a statistically significant benefit to estimating the partition structure as opposed to equally dividing the domain. Additionally, in Figure \ref{fig:bias}, we conclude that in all simulations settings, the mean slope estimate is unbiased in simulation as the true value is contained in the confidence intervals. Hence, we conclude our estimation procedure demonstrates little sacrifice in accuracy, no bias in mean parameter estimation, and therefore this method can be leveraged to analyze our data. 

\section{Data Analysis}
\label{data}
We use the model and methods presented in this article to provide inference on the spatial correlation of arsenic's bonding behavior using the data described in the Introduction. Because of the heterogeneous elemental composition of sand grains, we expect arsenic to accumulate on the sand grain differently by spatial location. We hypothesize these differences can be characterized by the mean and covariance of accumulation. Since the elemental composition changes sharply over the space, we expect these non-stationarities to be potentially sharp and varied. However, the non-stationarity of arsenic's bonds has not been well studied. Therefore, our primary objective in this analysis is to first evaluate whether the accumulation of arsenic is better modeled by a non-stationary model through implementation of our partition selection method. We then want to characterize the differences in estimated mean and covariance parameters in these blocks to draw inference that could motivate further directed research in this area.

The data were collected at the Submicron Resolution X-ray Spectroscopy (SRX) beamline at the NSLS-II synchrotron Brookhaven National Laboratory. This newly-developing technology was designed for high quality sub-micron spectroscopy, where complex heterogeneous elemental compositions can be identified \citep{andrade2011submicron}. We used step scanning as a means of collection of flourescence data with a dwell time of 0.5 seconds. Using a very high intensity beam at 14 keV, above the absorption edge of arsenic of 11.8667 keV,  the relative elemental abundances and speciation of arsenic of a (sub)micrometer area of a sand grain were measured. A detector positioned close to the sample recorded the count of fluoresced photons returned from the irradiated portion of the sample at various known frequencies, which allowed us to measure the abundance of other elements such as iron, calcium, and titanium simultaneously. This technology allows us to form elemental 100x100 $\mu m^2$ image maps at a about a 0.5 micrometer overlapping pixel resolution (n=40,000) for the various elements. The images obtained consisted primarily of the mineral and oxide outer coating of a principally quartz sand grain. 
\begin{figure}
	\begin{center}
		\includegraphics[width=5.5in]{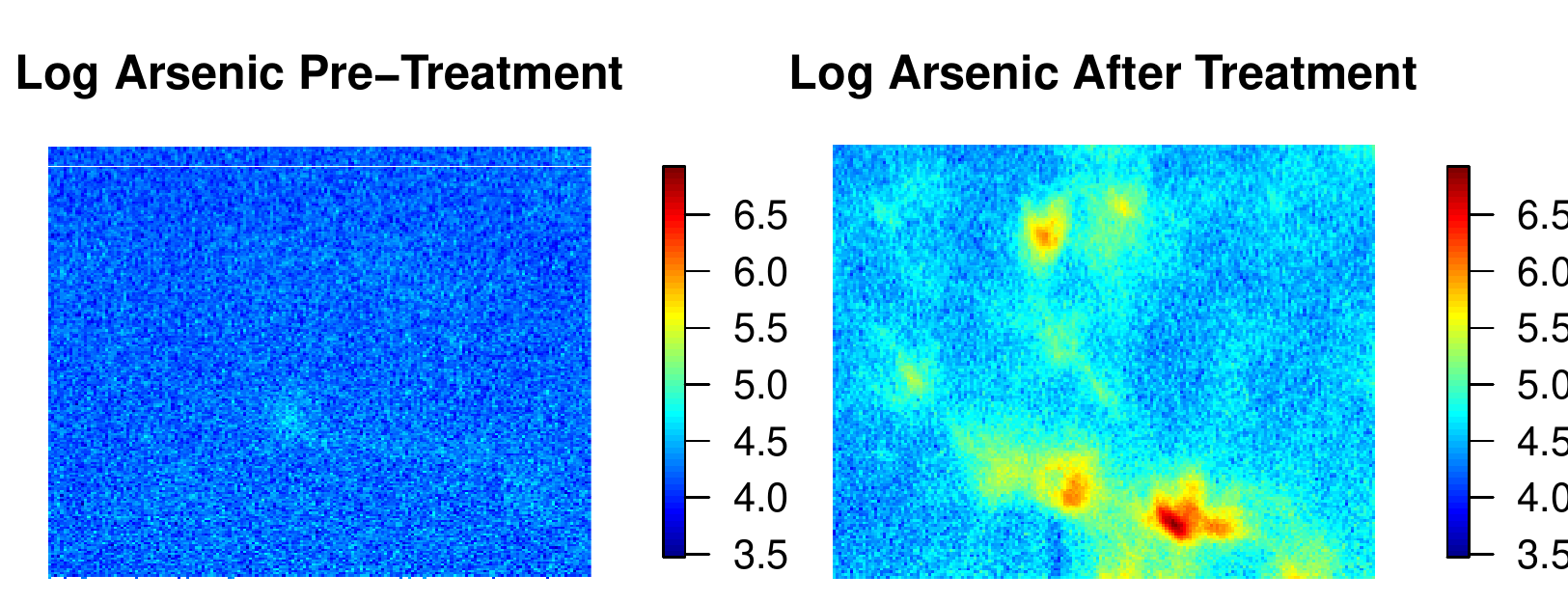}
	\end{center}
	\caption{Arsenic flourescence before and after lab treatment.}
	\label{fig:prepost}
\end{figure}

The sand grain is first mapped to understand pre-existing elemental structures. Then the sample is treated with sodium arsenate and is re-mapped at the same location. Because the sample was removed from the beam for treatment and then replaced, some realignment is necessary. Hence, we are left with a 180 x 150 pixel data matrix. In Figure \ref{fig:prepost} we see there is no signal (beyond scattering noise) of any arsenic accumulation prior to treatment. However, once the treatment is performed, a distinctive possibly non-stationary spatial pattern of arsenic is present.
\begin{figure}
	\begin{center}
		\includegraphics[width=4.5in]{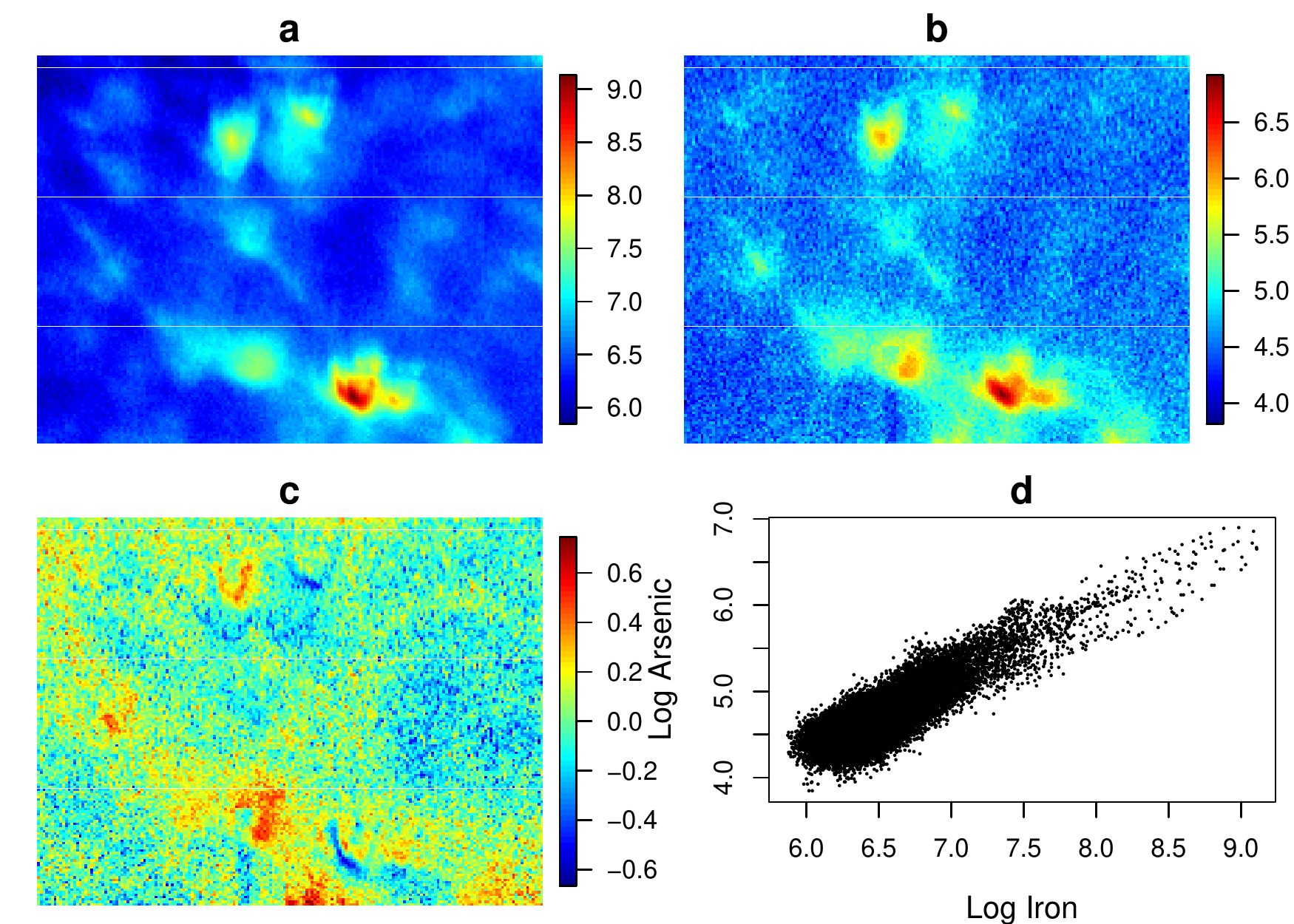}
	\end{center}
	\caption{a. Log iron flourescence before treatment. b. Log Arsenic flourescence after treatement. c. Residuals of linear fit of simple linear regression of a. and b. d. Scatterplot of a. and b.}
	\label{fig:four}
\end{figure}

\cite{guinness2014multivariate} noted a strong arsenic-iron correlation, and therefore we use the pre-treatment iron concentration map as a linear predictor of arsenic accumulation. With a log transformation of both variables, the linear fit seems plausible. The residuals from this fit seem to exhibit changing spatial covariances over the area. For example, the hotspot in the residuals map seems to exhibit high spatial covariance for low distances, but most of the map shows low spatial covariance for adjacent points. 

\begin{figure}
	\begin{center}
		\includegraphics[width=5.3in]{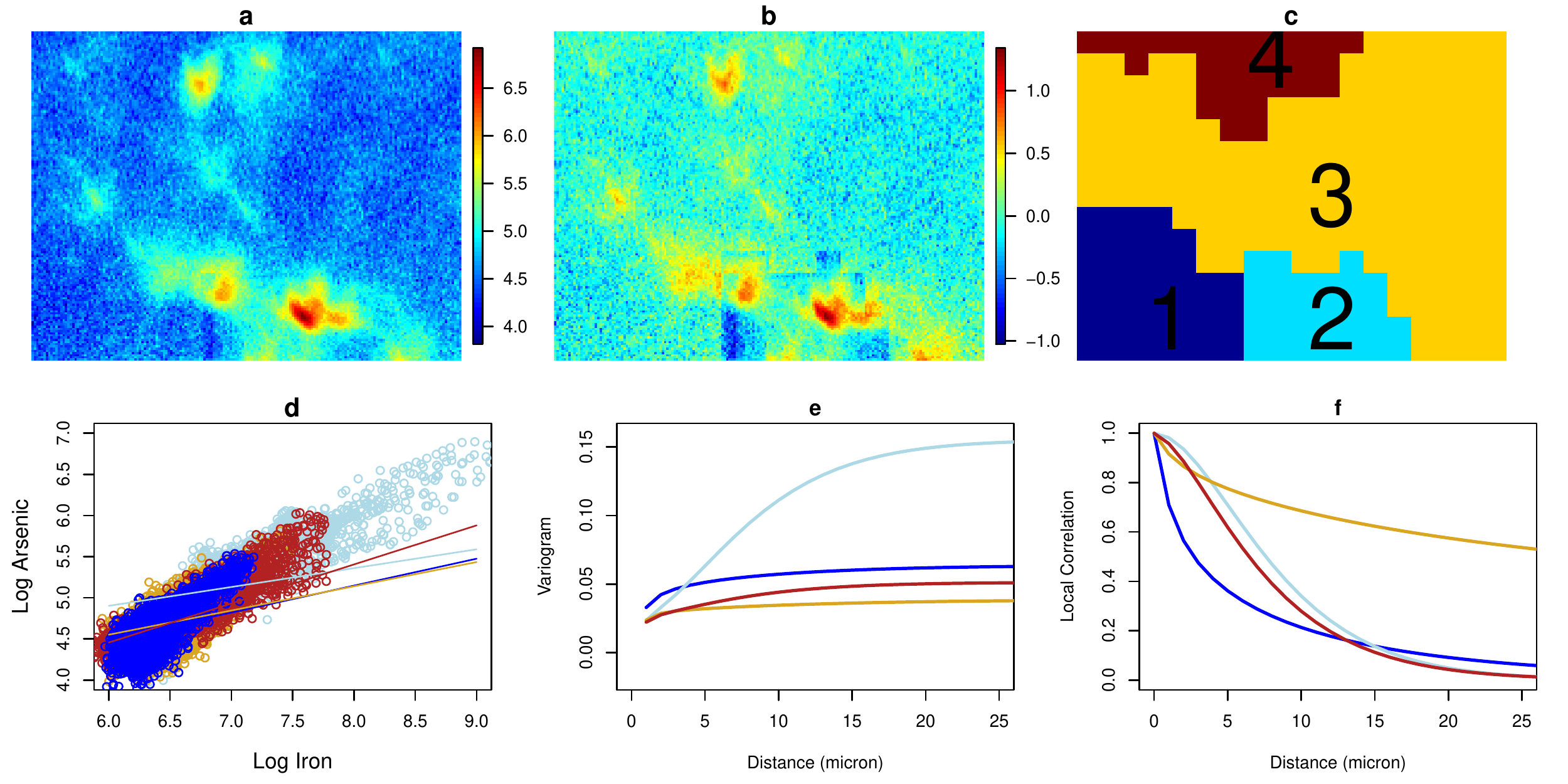}
	\end{center}
	\caption{a.\ Log arsenic distribution. b.\ Spatial residuals recovered a different regression in each block. c.\ Selected partition structure for analysis (with color scheme that matches the subsequent plots). d.\ Maximized parameter estimates of mean influence of log iron on log arsenic. e.\ Maximized variograms where black is the overall spectrum variogram. f.\ Maximized local correlation functions.}
	\label{fig:six}
\end{figure}

We use the the residuals from the simple linear regression fit in order to select a pool of 100 possible partitions. The BIC criterion discussed in Section \ref{sec:partition} returned the partition shown in \ref{fig:six} (c). Then we compare the BIC for these maximized surrogate partitions and select the partition structure produced by p-value = 0.001. This model can be seen in the below partition structure and is then assumed as fixed $\omega_i$ for maximum likelihood estimation. We formally define the model we estimate as:

$$log(As(s)) =  Z_{0}(s) + \sum_{i=1}^4 \omega_i(s) Z_{i}(s)$$
where  $s$ is the spatial location and
$Z_{i} \sim GP(\mu_i, C_i)$
$$ \mu_i(s) =\beta_{0i} + \beta_{1i} log(Fe(s)) $$
For identifiability, we assume $\mu_0=0$. Also, we assume that $C_i$ is quasi-Mat\'ern with a nugget. In applying the stochastic score approximation, we use 5 $U_j$ vectors.

The estimates of $\beta_{0i}$ and $\beta_{1i}$ exhibit a diverse set of relationships between log arsenic and log iron. By accounting for the complex non-stationarity, the mean influence of log iron on log arsenic accumulation is much smaller than that estimated via simple linear regression. The largest slope estimate is given in block 4, which encompasses one of areas of the hotspots of iron accumulation. The opposite is true in block 2, where there is the highest arsenic accumulation yields the lowest slope estimate. All slope estimates are statistically different, but the estimates for blocks 1 and 3 are qualitatively similar. One possible explanation would be that there is complex multivariate interactions with other elements in these hotspots unused in this analysis that cause the varied accumulation behavior. Another cause could be the influence of varying topography of the sand grain on binding architecture. Further research may be done in order to determine the underlying cause.

\begin{table}[ht]
	\centering
	\begin{tabular}{|c|cc|cccc|}
		\hline
		& $\beta_0$ & $\beta_1$ & $\sigma^2$ & $\rho$ & $\nu$ & $\tau$ \\ 
		\hline
		$Z_0$ & --  & --  & 0.013 & 0.586 & 0.032 & 0.006 \\ 
		$Z_1$ & 2.528 & 0.327 & 1.602 & 15.791 & 0.069 & 0.005 \\ 
		$Z_2$ & 3.528 & 0.229 & 5.151 & 3.581 & 2.176 & 0.005 \\ 
		$Z_3$ & 2.785 & 0.294 & 1.182 & 46.012 & 0.204 & 0.005 \\ 
		$Z_4$ & 1.607 & 0.475 & 1.023 & 4.304 & 1.313 & 0.005 \\ 
		\hline
	\end{tabular}
	\caption{Estimated mean and covariance parameters in the final data analysis model.}
	\label{tab:pars}
\end{table}

With selected parameter estimates, we see that the contribution of the overall stationary portion of the covariance structure is relatively small. Since the model is weighted in favor of the non-stationary portion of the model, we interpret the parameters as a means to dampen the edge effects of drops in the correlation for points on the edge of the blocks. In contrast, the locally stationary portions of the model contribute highly to the ultimate covariance defined by the model. The blocks 2 and 4 have similar spatially smooth correlation structure with large smoothness parameter estimates and moderate range estimates, but block 2 has a significantly higher variance than block 4. Blocks 1 and 3 both demonstrate less correlation at short distances, but block 3 demonstrates lingering long-scale correlation. Although the mean parameters were relatively similar for blocks 1 and 3, their correlation functions vary significantly, highlighting the necessity of fitting a non-stationary covariance model. In conclusion, the blocks that contain hotspots of arsenic have different mean accumulation patterns, but similar spatially smooth covariance functions.

\section{Conclusion}
\label{sec:conc}

In this paper, we have presented a new estimation method for a non-stationarity model represented as the weighted linear combination of locally stationary processes and a globally stationary process. We estimate the parameters in large gridded datasets with only $O(n \log n)$ computational complexity per each step of estimation and $O(n)$ storage. The biggest hurdle in such a model is often the definitions of the locally stationary processes, and we offer an algorithm that uses the data to estimate irregularly shaped non-stationary processes. We demonstrate that using BIC to choose between candidate partitions offers a significant advantage in increasing the Rand index of a partition, therefore better approximating the underlying non-stationary process. 

Second, we introduced a method of parameter estimation that generalized the use of the stochastic score approximation and other computational tools to the non-stationary case.  Within our estimation method, we proposed a pre-conditioner matrix that approximates the inverse of the induced covariance matrix without the need to form the large pre-conditioner matrix, which reduces computing time by approximately 3 times under any parameter estimates. We generalize the use circulant embedding to this non-stationary case, and describe implementation of the complex estimation algorithm where step-size selection is non-trivial.

In simulation, we have showed that our estimation method is more accurate and up to 24 times faster than competing methods even in relatively small sample sizes even with only one approximation vector. Finally, by applying our method to x-ray fluorescence data, we were able to see that arsenic has significant non-stationarity in both the mean and covariance parameter estimates. Spatially smooth covariance structure in the hotspots of arsenic unexplained by iron accumulation motivates further scientific inquiry. 

 This estimation method could be adapted to a multivariate or spatio-temporal framework which would involve further work on defining multivariate or spatial-temporal non-stationary models and developing the corresponding computational tools to evaluate the stochastic score efficiently. Alternately, many large datasets have irregularly spaced observations. Further work could be done in order to extend these methods to the non-gridded case where an approximate fine grid is utilized as a surrogate to the irregular locations.
Additionally, the uncertainty quantification of the parameters does not account for the selected partition structure. Bayesian model averaging may be able to be applied to fully specify the posterior distributions of the parameters over these partitions.
Additionally, Bayesian estimation methods would offer a natural framework with which to consider prediction and kriging from the model.

In conclusion, we have presented a non-stationary model that is scientifically intuitive and can account for sharp changes in spatial correlation across the field. We have found that stochastic approximation to the score function is fast at estimating the maximum likelihood parameters in the gridded data case. It offers little loss in likelihood maximization when evaluating changes in spatial correlation across the field and makes evaluation of a complex non-stationary model viable for large data sets.

\bibliographystyle{Chicago}
\bibliography{bib}

\begin{thebibliography}{}

\bibitem[\protect\citeauthoryear{Andrade, Thieme, Northrup, Yao, Lanzirotti,
  Eng, and Shen}{Andrade et~al.}{2011}]{andrade2011submicron}
Andrade, V.~D., J.~Thieme, P.~Northrup, Y.~Yao, A.~Lanzirotti, P.~Eng, and
  Q.~Shen (2011).
\newblock The sub-micron resolution x-ray spectroscopy beamline at nsls-ii.
\newblock {\em Nuclear Instruments and Methods in Physics Research Section A:
  Accelerators, Spectrometers, Detectors and Associated Equipment\/}~{\em
  649\/}(1), 46 -- 48.
\newblock National Synchrotron Radiation Instrumentation conference in 2010.

\bibitem[\protect\citeauthoryear{Anitescu, Chen, and Wang}{Anitescu
  et~al.}{2012}]{anitescu2012matrix}
Anitescu, M., J.~Chen, and L.~Wang (2012).
\newblock A matrix-free approach for solving the parametric gaussian process
  maximum likelihood problem.
\newblock {\em SIAM Journal on Scientific Computing\/}~{\em 34\/}(1),
  A240--A262.

\bibitem[\protect\citeauthoryear{Chen, Wang, and Anitescu}{Chen
  et~al.}{2014}]{chen2014fast}
Chen, J., L.~Wang, and M.~Anitescu (2014).
\newblock A fast summation tree code for matérn kernel.
\newblock {\em SIAM Journal on Scientific Computing\/}~{\em 36\/}(1),
  A289--A309.

\bibitem[\protect\citeauthoryear{Dennis~Jr and Schnabel}{Dennis~Jr and
  Schnabel}{1996}]{dennis1996numerical}
Dennis~Jr, J.~E. and R.~B. Schnabel (1996).
\newblock {\em Numerical methods for unconstrained optimization and nonlinear
  equations}, Volume~16.
\newblock Siam.

\bibitem[\protect\citeauthoryear{Fuentes}{Fuentes}{2001}]{fuentes2001high}
Fuentes, M. (2001).
\newblock A high frequency kriging approach for non-stationary environmental
  processes.
\newblock {\em Environmetrics\/}~{\em 12\/}(5), 469--483.

\bibitem[\protect\citeauthoryear{Fuentes}{Fuentes}{2002}]{fuentes2002interpolation}
Fuentes, M. (2002).
\newblock Interpolation of nonstationary air pollution processes: a spatial
  spectral approach.
\newblock {\em Statistical Modelling\/}~{\em 2\/}(4), 281--298.

\bibitem[\protect\citeauthoryear{Fuentes, Chaudhuri, and Holland}{Fuentes
  et~al.}{2007}]{fuentes2007bayesian}
Fuentes, M., A.~Chaudhuri, and D.~M. Holland (2007).
\newblock Bayesian entropy for spatial sampling design of environmental data.
\newblock {\em Environmental and Ecological Statistics\/}~{\em 14\/}(3),
  323--340.

\bibitem[\protect\citeauthoryear{Guinness and Fuentes}{Guinness and
  Fuentes}{2015}]{guinness2015likelihood}
Guinness, J. and M.~Fuentes (2015).
\newblock Likelihood approximations for big nonstationary spatial temporal
  lattice data.
\newblock {\em Statistica Sinica\/}~{\em 25\/}(1), 329--349.

\bibitem[\protect\citeauthoryear{Guinness and Fuentes}{Guinness and
  Fuentes}{2017}]{guinness2017circulant}
Guinness, J. and M.~Fuentes (2017).
\newblock Circulant embedding of approximate covariances for inference from
  gaussian data on large lattices.
\newblock {\em Journal of Computational and Graphical Statistics\/}~{\em
  26\/}(1), 88--97.

\bibitem[\protect\citeauthoryear{Guinness, Fuentes, Hesterberg, and
  Polizzotto}{Guinness et~al.}{2014}]{guinness2014multivariate}
Guinness, J., M.~Fuentes, D.~Hesterberg, and M.~Polizzotto (2014).
\newblock Multivariate spatial modeling of conditional dependence in microscale
  soil elemental composition data.
\newblock {\em Spatial Statistics\/}~{\em 9}, 93--108.

\bibitem[\protect\citeauthoryear{Guinness and Stein}{Guinness and
  Stein}{2013}]{guinness2013transformation}
Guinness, J. and M.~L. Stein (2013).
\newblock Transformation to approximate independence for locally stationary
  gaussian processes.
\newblock {\em Journal of Time Series Analysis\/}~{\em 34\/}(5), 574--590.

\bibitem[\protect\citeauthoryear{Haas}{Haas}{1995}]{haas1995local}
Haas, T.~C. (1995).
\newblock Local prediction of a spatio-temporal process with an application to
  wet sulfate deposition.
\newblock {\em Journal of the American Statistical Association\/}~{\em
  90\/}(432), 1189--1199.

\bibitem[\protect\citeauthoryear{Higdon}{Higdon}{1998}]{Higdon1998aprocess}
Higdon, D. (1998, Jun).
\newblock A process-convolution approach to modelling temperatures in the north
  atlantic ocean.
\newblock {\em Environmental and Ecological Statistics\/}~{\em 5\/}(2),
  173--190.

\bibitem[\protect\citeauthoryear{Higdon, Swall, and Kern}{Higdon
  et~al.}{1999}]{higdon1999nonstationary}
Higdon, D., J.~Swall, and J.~Kern (1999).
\newblock Non-stationary spatial modeling.
\newblock {\em Bayesian statistics\/}~{\em 6\/}(1), 761--768.

\bibitem[\protect\citeauthoryear{Hutchinson}{Hutchinson}{1990}]{hutchinson1990stochastic}
Hutchinson, M.~F. (1990).
\newblock A stochastic estimator of the trace of the influence matrix for
  laplacian smoothing splines.
\newblock {\em Communications in Statistics-Simulation and Computation\/}~{\em
  19\/}(2), 433--450.

\bibitem[\protect\citeauthoryear{Nesterov}{Nesterov}{2005}]{nesterov2005smooth}
Nesterov, Y. (2005).
\newblock Smooth minimization of non-smooth functions.
\newblock {\em Mathematical programming\/}~{\em 103\/}(1), 127--152.

\bibitem[\protect\citeauthoryear{Rand}{Rand}{1971}]{rand1971objective}
Rand, W.~M. (1971).
\newblock Objective criteria for the evaluation of clustering methods.
\newblock {\em Journal of the American Statistical association\/}~{\em
  66\/}(336), 846--850.

\bibitem[\protect\citeauthoryear{Risser, Calder, Berrocal, and Berrett}{Risser
  et~al.}{2016}]{risser2016nonstationary}
Risser, M.~D., C.~A. Calder, V.~J. Berrocal, and C.~Berrett (2016).
\newblock Nonstationary spatial process modeling via treed covariate
  segmentation, with application to soil organic carbon stock assessment.
\newblock {\em arXiv preprint arXiv:1608.05655\/}.

\bibitem[\protect\citeauthoryear{Sampson and Guttorp}{Sampson and
  Guttorp}{1992}]{sampson1992nonparametric}
Sampson, P.~D. and P.~Guttorp (1992).
\newblock Nonparametric estimation of nonstationary spatial covariance
  structure.
\newblock {\em Journal of the American Statistical Association\/}~{\em
  87\/}(417), 108--119.

\bibitem[\protect\citeauthoryear{Stein, Chen, Anitescu, et~al.}{Stein
  et~al.}{2013}]{stein2013stochastic}
Stein, M.~L., J.~Chen, M.~Anitescu, et~al. (2013).
\newblock Stochastic approximation of score functions for gaussian processes.
\newblock {\em The Annals of Applied Statistics\/}~{\em 7\/}(2), 1162--1191.

\bibitem[\protect\citeauthoryear{Vecchia}{Vecchia}{1988}]{vecchia1988estimation}
Vecchia, A.~V. (1988).
\newblock Estimation and model identification for continuous spatial processes.
\newblock {\em Journal of the Royal Statistical Society. Series B
  (Methodological)\/}, 297--312.

\end{thebibliography}

\section{Acknowledgments}

The authors gratefully acknowledge the contribution of Dr. Dean Hesterberg and Aakriti Sharma for their contributions to the project through scientific collaboration and supplying the data. 

This material is based upon work supported by NSF Research Network on Statistics in the Atmosphere and Ocean Sciences (STATMOS) through grants DMS-1106862 and DMS-1107046 as well as NSF-DMS Grant Numbers 1406016, 1613219, and 1723158. It was also funded partially through NSF grant 570235.

Research reported in this publication was supported by the National Institutes of Health under award number R01ES027892.

This material was based upon work partially supported by the National Science Foundation under Grant DMS-1638521 to the Statistical and Applied Mathematical Sciences Institute. Any opinions, findings, and conclusions or recommendations expressed in this material are those of the author(s) and do not necessarily reflect the views of the National Science Foundation.

This research used resources of the SRX beamline at the National Synchrotron Light Source II, a U.S. Department of Energy (DOE) Office of Science User Facility operated for the DOE Office of Science by Brookhaven National Laboratory under Contract No. DE-SC0012704.

\newpage
\begin{center}
{\large\bf SUPPLEMENTARY MATERIALS}
\end{center}
\section{Appendix A}
Below are the results of a simulation where we compare the absolute error in the approximation to the score when use of the dependent sampling scheme developed by \cite{stein2013stochastic} and the original stochastic score \citep{hutchinson1990stochastic}. There is a significant benefit to the dependent sampling scheme when the number of vectors are large, but since we show very few vectors are needed for approximate solutions, we adpot \cite{hutchinson1990stochastic}.

\begin{figure}[h]
	\begin{center}
		\includegraphics[width=4in]{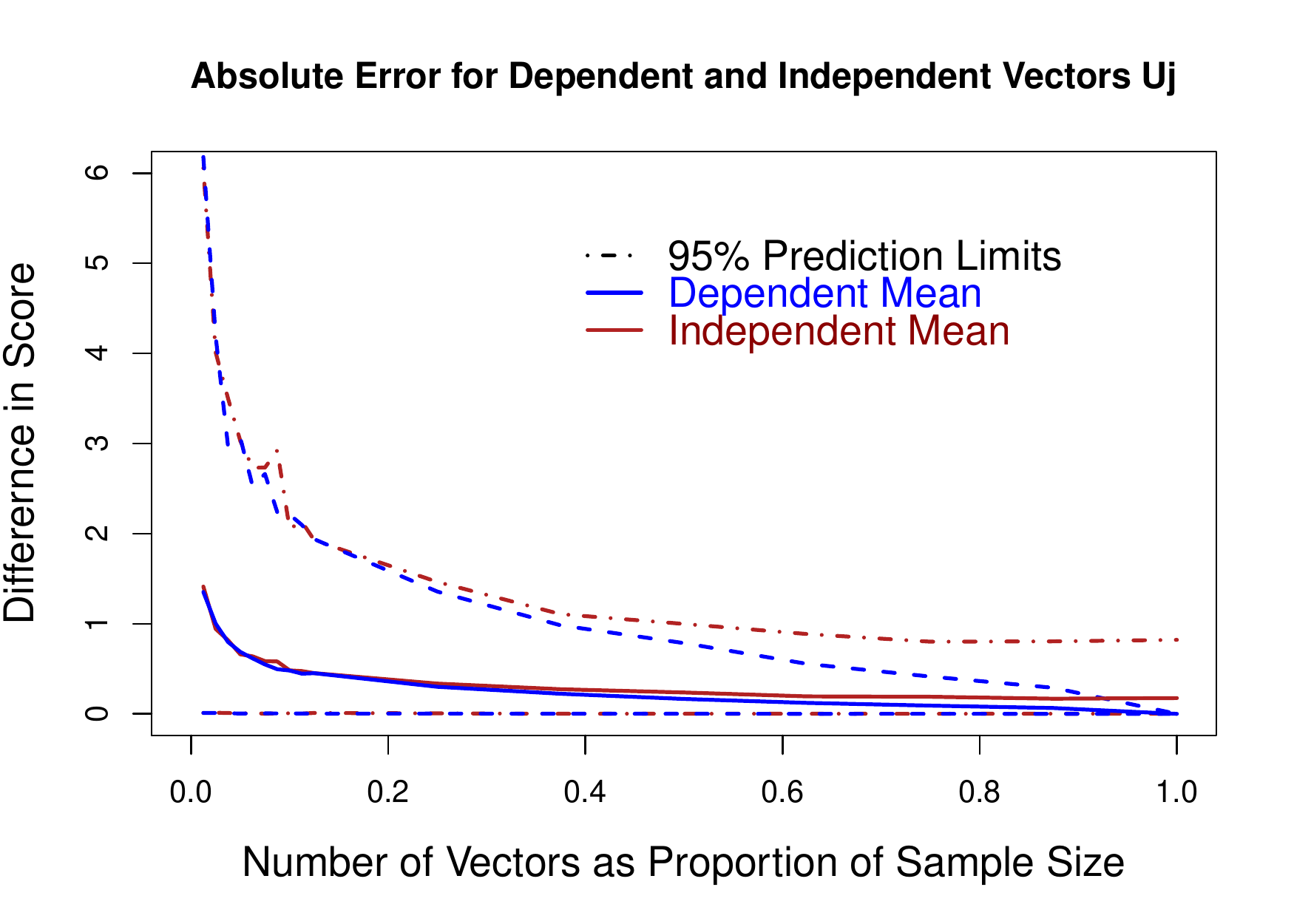}
	\end{center}
	\caption{Absolute Error in estimating the score using the independent and dependent generation schemes. The dotted lines represent 95\% prediction intervals for the absolute error in score estimation using 100 iterations at each sample size. The total sample size for the simulation was 400. }
\end{figure}

\section{Appendix B}
For the stationary quasi-Mat\'ern spectral density, we define the normalizing constant $c$ so that
$$ c = \frac{1}{\sum_{\gamma} \{\alpha^2 + \sin^2(\frac{\gamma_1}{2})+\sin^2(\frac{\gamma_2}{2})\}^{(-1-\nu)}} $$

For ease of notation, let $g = \{\alpha^2 + \sin^2(\frac{\gamma_1}{2})+\sin^2(\frac{\gamma_2}{2})\}^{(-1-\nu)}$ and thus $ c = \frac{1}{\sum_{\gamma} g} $.

Then the quasi-Mat\'ern spectral density is

$$f(\gamma)  = \sigma^2 [\frac{1}{\sum_{\gamma}  g } (1-\tau) \{\alpha^2 + [\sin^2(\frac{\gamma_1}{2}) + \sin^2(\frac{\gamma_2}{2})]\}^{-1-\nu} + \tau]$$
 
 Therefore the differentiated spectral densities are as follows:
 $$\frac{\partial}{\partial \sigma^2} = \frac{ f(\gamma)}{\sigma^2}$$

  $$\frac{\partial}{\partial \alpha^2}  = 2 \alpha (1+\nu) \sigma^2 c (1-\tau) \{1 - gc\} g^{1 + \frac{1}{\nu+1}}$$

  $$\frac{\partial}{\partial \nu} = \sigma^2 c (1-\tau) g \{1 - gc\}  log(g^{-\frac{1}{\nu+1}})$$

 $$\frac{\partial}{\partial \tau}   = \sigma^2 \{1 - gc\} $$
 
 For optimization purposes, we suggest re-parameterizing $\sigma^2, \alpha, \nu$ on the log scale and $\tau$ on the logit scale.

\section{Appendix C}
\citet{stein2013stochastic} show this formulation is convenient computationally in the stationary case since we can avoid forming the covariance matrix directly through iterative solves and straightforward circulant embedding. We extend this work by proposing a non-stationary model that can leverage adaptations of these tools in this more complex case. Thus, we are able to evaluate an unbiased approximation of the score in $O(n)$ storage and $O(n\log(n))$ operations as compared to $O(n^2)$ storage and $O(n^3)$ operations of computing the likelihood directly for a full-rank, non-stationary covariance matrix. 

By computing the Fisher information matrix for $\theta$, the asymptotic variance of the maximum likelihood parameters in the multivariate normal are $\frac{1}{2}tr(K^{-1}(\theta) K_i(\theta)K^{-1}(\theta) K_i(\theta))$ (Stein, 1999). Therefore, in Stein et.\ al (2013), they derive that the $i,j$ element of the information matrix $B$ of estimates obtained via stochastic score have elements:
$$B_{i,j} =(\frac{1}{2} + \frac{1}{2N})tr(K^{-1}(\theta) K_i(\theta)K^{-1}(\theta) K_j(\theta)) - \frac{1}{2N} tr((K_i K^{-1}) * (K_j K^{-1}))$$

where $*$ is element-wise multiplication. In this formulation, we observe a few properties of the approximation to keep in mind. First, as $N \to \infty$, the asymptotic variance of the estimates approaches that of the traditional maximum likelihood estimation. Additionally, it is important to observe that the variance of the approximation does vary by the function of the parameter and therefore parameters with large maximum likelihood variances, will also have higher variance induced by use of the stochastic score.

\section{Appendix D}
\begin{figure}
	\begin{center}
		\includegraphics[width=5.5in]{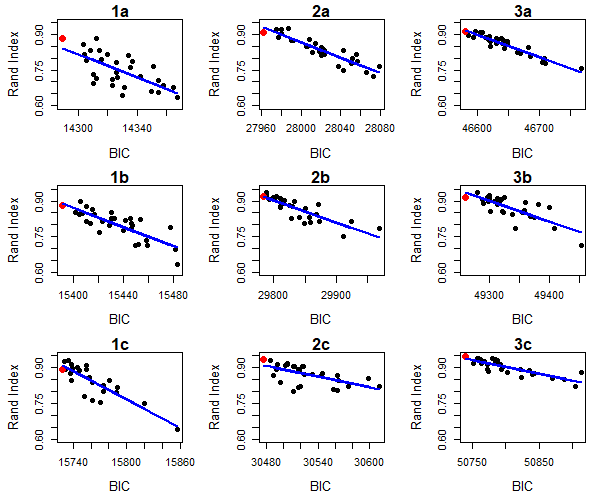}
	\end{center}
	\caption{Each plot shows one iteration of partition estimation for sample size (1,2,3) and non-stationary setting (a,b,c). All regression lines are statistically significant. Red point indicates the chosen partition in each case.}
	\label{fig:bic}
\end{figure}

As you can see from Figure \ref{fig:bic}, in all settings, there is a statistically significantly negative correlation between BIC and Rand index. Since we cannot calculate the Rand index of an unknown partition structure, this validates the use of BIC as a selection criterion.

\section{Appendix E}
In this section, we outline the non-linear solver in Section \ref{opt}. Since likelihood evaluation is not possible in this case, we define an algorithm for step size selection so that the direction of the gradient is not changed by parameter movement. Hence, if we assume our starting value is in a convex space that contains the solution, we do not step over that solution, and therefore find it more quickly. We first select the stochastic score vectors $U_j$ and fix these throughout the algorithm. As a starting value, we assume that the parameter settings are the same in all blocks, and are therefore stationary. Let  $g = (g_1, g_2, ... g_p)$ be the approximate score at these initial parameters and $g_0 = max(|g_i|)$. As an initial step size choice, we have found that $\frac{.5}{g_0}$ works well. Additionally, we initialize a threshold for movement as $\frac{g_0}{10}$. By moving only the parameters with absolute value scores larger than this threshold, we have found we are able to move toward the solution more quickly by selecting a larger step size. 
Then, our modified proximal gradient algorithm has the following steps in each iteration:
\begin{enumerate}
	\item For all parameters with an absolute score value above the threshold (excluding $\sigma^2_0$ and $\beta$), we propose new parameters as $$\text{candidate parameter} = \text{parameter} + \text{step size}*\text{approximate score}$$
	\item Solve for $\sigma_0^2$ in the candidate parameters as shown in Section \ref{opt}
	\item Project candidate parameters into space where there are no computational difficulties with singularity of the covariance matrix.
	\item Evaluate approximate score of these candidate parameters.
	\item Test if new approximate score has the same sign as current parameter's approximate score for each parameter.
	\begin{enumerate}
		\item If it is the same all cases, we update parameters as the candidate parameters and solve for $\beta$ and $\sigma^2_0$. Again, we project the covariance parameters into a computationally feasible space. Then, we evaluate the approximate score for direction in the next step. Step size is multiplied by 2, and movement tolerance for the next step is re-evalutated as $\frac{max(|g_i|)}{10}$, where $g = (g_1, g_2, ... g_p)$ is the current approximate score.
		\item Otherwise, step size is multiplied by 0.1 and the candidate parameters are rejected.
	\end{enumerate}
\end{enumerate}

As a stopping criterion, we exit the algorithm when the maximum absolute value of approximate score value is less than $\frac{2*g_0}{\sqrt n}$.

\end{document}